\documentclass[12pt]{article}

\usepackage{amssymb}
\usepackage{amsthm}

\usepackage{logique}

\usepackage{graphicx}
\usepackage[euler]{textgreek}

\renewcommand{\force}{\;\|\hspace{-0.5em}-}

\newcommand{\NN}{\mathbb{N}}

\newcommand{\III}{\mbox{\sffamily I}}

\newcommand{\OOO}{\mbox{\sffamily O}}

\newcommand{\RRR}{\mbox{\sffamily R}}

\newcommand{\sbs}{\sqsubset}
\newcommand{\ssbs}{\sbs\hspace{-0.65em}\sbs}

\newcommand{\et}{{\scriptstyle\land}}
\newcommand{\ou}{{\scriptstyle\lor}}
\newcommand{\non}{{\scriptstyle\neg}}

\newcommand{\ZFe}{ZF$_\varepsilon$}

\newcommand{\lbr}{\langle}
\newcommand{\rbr}{\rangle}
\newcommand{\gl}{\gimel}
\newcommand{\hto}{\hookrightarrow}

\newtheorem{theorem}{Theorem}
\newtheorem{lemma}[theorem]{Lemma}
\newtheorem{corollary}[theorem]{Corollary}
\newtheorem{proposition}[theorem]{Proposition}

\author{Jean-Louis Krivine\\
\footnotesize{P.P.S. group,}
\footnotesize{University Paris-Diderot, CNRS}\\
\footnotesize{krivine@pps.univ-paris-diderot.fr}
}

\title{On the structure\\
of classical realizability models of ZF}
\date{\footnotesize \today}

\begin{document}
\maketitle\noindent
{\small{\bfseries Abstract.} In \cite{krivine5,krivine6,krivine7}, we have introduced the technique of
\emph{classical realizability}, which permits to extend the Curry-Howard correspondence between proofs
and programs, to Zermelo-Fraenkel set theory. The models of ZF we obtain in this way, are called
\emph{realizability models}~; this technique is an extension of the method of forcing,
in which the ordered sets (sets of \emph{conditions}) are replaced with more complex first order
structures called \emph{realizability algebras}.\\
We show here that every realizability model ${\cal N}$ of ZF contains a transitive submodel, which has
the same ordinals as ${\cal N}$, and which is an elementary extension of the ground model. It follows
that the constructible universe of a realizability model is an elementary extension of the constructible
universe of the ground model.\\
We obtain this result by showing the existence of an ultrafilter on the \emph{characteristic
Boolean algebra}~$\gl2$ of the realizability model, which is defined in~\cite{krivine6,krivine7}.	}

\section*{Introduction}\noindent
We use here the basic notions and notations of the theory of \emph{classical realizability},
which was developed in \cite{krivine5,krivine6,krivine7}.\\
We consider a model ${\cal M}$ of \ ZF + V = L,  which we call the \emph{ground model}
\footnote{In fact, it suffices that ${\cal M}$ satisfy the \emph{choice principle CP}, which is written
as follows, in the language of ZF with a new binary relation symbol $\triangleleft$~:
{\em ``~$\triangleleft$ \ is a well ordering relation on ${\cal M}$''.}\\
It is well known  that, in every countable model of ZFC, we can define such a binary symbol, so as to
get a model of ZF~+~CP\/. Thus, ZF~+~CP is a \emph{conservative extension} of ZFC.}\nopagebreak
\ and, in~${\cal M}$, a \emph{realizability algebra} ${\cal A}=(\Lbd,\Pi,\Lbd\star\Pi,\mbox{QP},\bbot)$.\\
$\Lbd$ is the set of \emph{terms}, $\Pi$ is the set of \emph{stacks}, $\Lbd\star\Pi$ is the set of
\emph{processes}, $\mbox{QP}\subset\Lbd$ is the set of \emph{proof-like terms}, and $\bbot$ is a
distinguished subset of $\Lbd\star\Pi$.\\
They satisfy the axioms of \emph{realizability algebra}, which are given in \cite{{krivine5}} or \cite{krivine7}.\\
In the model ${\cal M}$, we use the language of ZF with the binary relation symbols $\notin,\subset$
and function symbols, which we shall define when needed, by means of formulas of ZF\/.\\
We can now build (see~\cite{krivine5}) the \emph{realizability model} ${\cal N}$, which has the same set of
individuals as ${\cal M}$, the truth value set of which is ${\cal P}(\Pi)$, endowed with a suitable Boolean algebra
structure.\\
The language of this model has three binary relation symbols $\neps,\notin,\subset$, and the same function symbols
as the model ${\cal M}$, with the same interpretation.

\smallskip\noindent
The \emph{formulas} are built as usual, from atomic formulas, \emph{with the only logical
symbols $\bot,\to,\pt$}. We shall use the notations~:\\
$\neg F$ for $F\to\bot$~; $F_1,\ldots,F_n\to F$ for $(F_1\to(\ldots\to F_n)\to F$~;\\
$\ex x\,F$ for $\neg\pt x\neg F$~; $\ex x\{F_1,\ldots,F_n\}$ for $\neg\pt x(F_1,\ldots,F_n\to\bot)$.

\smallskip\noindent
{\small{\bfseries Notation.} We shall often use the notation $\vec{x}$ for a finite sequence
$x_1,\ldots,x_n$~; for instance, we shall write $F[\vec{x}]$ for $F[x_1,\ldots,x_n]$.}

\smallskip\noindent
By means of the completeness theorem, we obtain from ${\cal N}$ an ordinary model ${\cal N}'$,
with truth values in $\{0,1\}$. The set of individuals of ${\cal N}'$ generally
\emph{strictly contains} ${\cal N}$.\\
The elements of ${\cal N}'$ (resp. ${\cal M}$) are called \emph{individuals of} ${\cal N}$ (resp. ${\cal M}$).
The individuals are generally denoted by $a,b,c,\ldots,a_0,a_1,\ldots$

\smallskip\noindent
In \cite{krivine5} or \cite{krivine6}, we define a theory \ZFe, written in this language. We show that it is a
\emph{conservative extension} of ZF\/, and that the model ${\cal N}$ \emph{satisfies the axioms of \ZFe},
which means that each of these axioms is \emph{realized by a proof-like term}.\\
Given a term $\xi\in\Lbd$ and a closed formula $F[a_1,\ldots,a_n]$ in the language of \ZFe, with parameters
$a_1,\ldots,a_n$ in ${\cal N}$ (or, which is the same, in ${\cal M}$), we shall write~:\\
$\xi\force F[a_1,\ldots,a_n]$ in order to say that the term $\xi$ realizes $F[a_1,\ldots,a_n]$.\\
The truth value of this formula is a subset of $\Pi$, denoted by $\|F[a_1,\ldots,a_n]\|$.\\
We write $\force F[a_1,\ldots,a_n]$ in order to say that $F[a_1,\ldots,a_n]$ is realized by
some proof-like term.

\smallskip\noindent
Thus, the model ${\cal N}'$ satisfies \ZFe~; therefore, in ${\cal N}'$, we can define a model of ZF\/,
denoted~${\cal N}'_\in$, in which the equality is interpreted by \emph{the extensional equivalence}
denoted by $x\simeq y$ (that is ${x\subset y\land y\subset x}$).

\smallskip\noindent
The general properties of the realizability models are described in~\cite{krivine7}~; we shall use
the definitions and notations of this paper.

\smallskip\noindent
In what follows, unless otherwise stated, each formula of \ZFe\ must be interpreted in ${\cal N}$
(its truth value is a subset of $\Pi$) or, if one prefers, in ${\cal N}'$
(then its truth value is $0$ or $1$). If the formula must be interpreted in ${\cal M}$,
(in that case, it does not contains the symbol $\not\!\veps$) it will be explicitly stated.

\section*{Function symbols}\noindent
{\bfseries Notations.} The formula \ $\pt z(z\neps y\to z\neps x)$ \ is denoted by \ $x\subseteq y$
(\emph{strong inclusion})~;\\
the formula \ $x\subseteq y\land y\subseteq x$ \ is denoted by \ $x\cong y$
(\emph{strong extensional equivalence}).\\
We recall that $\subset$ and $\simeq$ are the symbols of inclusion and of extensional equivalence of~ZF~:\\
$x\subset y\equiv\pt z(z\notin y\to z\neps x)$~; \ $x\simeq y\equiv(x\subset y\land y\subset x)$.

\subsection*{Function symbols associated with axioms of \ZFe}\noindent
\smallskip\noindent
In this section, we define a function symbol for each of the following axioms of~\ZFe~:\\
comprehension, pairing, union, power set and collection.

\smallskip\noindent
{\bfseries Comprehension.}\\
For each formula $F[y,\vec{z}]$ of \ZFe\/, (where $\vec{z}$ is a finite sequence of variables $z_1,\ldots,z_n$)
we define, in ${\cal M}$, a symbol of function of arity $n+1$, denoted provisionally by
$\mbox{Compr}_F(x,\vec{z})$,
(Compr is an abbreviation for \emph{Comprehension}) by setting~:\\
$\mbox{Compr}_F(a,\vec{c})=\{(b,\xi\ps\pi)\;;\;(b,\pi)\in a,\;\xi\force F[b,\vec{c}]\}$.\\
It was shown in~\cite{krivine7} (and it is easily checked) that we have~:\\
$\|b\neps\mbox{Compr}_F(a,\vec{c})\|=\|F[b,\vec{c}]\to b\neps a\|$. Thus, we have~:\\
$\III\,\force\pt x\pt y\pt\vec{z}(y\neps\mbox{Compr}_F(x,\vec{z})\to(F[y,\vec{z}]\to y\neps x))$~;\\
$\III\,\force\pt x\pt y\pt\vec{z}((F[y,\vec{z}]\to y\neps x)\to y\neps\mbox{Compr}_F(x,\vec{z}))$.\\
Therefore, instead of $\mbox{Compr}_F(x,\vec{z})$, we shall use for this function symbol, the more intuitive
notation \ $\{y\eps x\;;\;F[y,\vec{z}]\}$, in which $y$ is a {\em bounded variable}.

\smallskip\noindent
{\bfseries Pairing.}\\
We define the following binary function symbol~:\\
\centerline{$\mbox{pair}(x,y)=\{z\eps\{x,y\}\fois\Pi\;;\;(z=x)\lor(z=y)\}$.}

\noindent
It is easily checked that we have the desired property~:\\
\centerline{$\force\pt x\pt y\pt z(z\eps\mbox{pair}(x,y)\dbfl z=x\lor z=y)$.}

\noindent
{\small{\bfseries{}Remark.} We could also define a symbol $\mbox{pair}(x,y)$, with this property,
directly in ${\cal M}$, as follows~:\\
\centerline{$\mbox{pair}(x,y)=\{(x,\ul{1}\ps\pi)\;;\;\pi\in\Pi\}\cup\{(y,\ul{0}\ps\pi)\;;\;\pi\in\Pi\}$.}}

\noindent
In the sequel, when working in ${\cal N}$, we shall use the (natural) abbreviations~:\\
$\{x,y\}$ for $\mbox{pair}(x,y)$~; $(x,y)$ for $\mbox{pair}(\mbox{pair}(x,x),\mbox{pair}(x,y))$.

\smallskip\noindent
{\bfseries Union and power set.}\\
We define below two unary function symbols
$\ov{\bigcup}x$ and $\ov{\cal P}(x)$, such that~:

\smallskip\noindent
$\force\pt x\pt z(z\eps\ov{\bigcup}x\dbfl(\ex y\eps x)\,z\eps y)$.

\smallskip\noindent
$\force\pt x(\pt y\eps\ov{\cal P}(x))(\pt z\eps y)(z\eps x)$~; \
$\force\pt x\pt y(\ex y'\eps\ov{\cal P}(x))\pt z(z\eps y'\dbfl z\eps x\land z\eps y)$.

\begin{theorem}\label{parties}
Let ${\cal V},{\cal Q}$ be the unary function symbols defined in ${\cal M}$ as follows~:\\
${\cal V}(a)=\mbox{Cl}(a)\fois\Pi$ \ and \ ${\cal Q}(a)={\cal P}(\mbox{Cl}(a)\fois\Pi)\fois\Pi$\\
where $\mbox{Cl}(a)$ is the transitive closure of $a$. Then, we have~:\\
i)~~$\III\,\force\pt x\pt y\pt z(z\eps y,z\neps{\cal V}(x)\to y\neps x)$.\\
ii)~~$\III\,\force\pt x\pt\vec{z}\left(\{y\eps x\;;\;F[y,\vec{z}]\}\eps{\cal Q}(x)\right)$ \
for every formula $F[x,\vec{z}]$ of \ZFe.
\end{theorem}\noindent
i)~~Let $a,b,c$ be individuals in ${\cal M}$, $\,\xi,\eta\in\Lbd$ and $\pi\in\Pi$ such that~:\\
$\xi\force c\eps b,\,\eta\force c\neps{\cal V}(a)$ and $\pi\in\|b\neps a\|$~; we have therefore $(b,\pi)\in a$.\\
We must show $\xi\star\eta\ps\pi\in\bbot$.\\
We show that $\|c\neps b\|\subset\|c\neps{\cal V}(a)\|$~: indeed, if $\rho\in\|c\neps b\|$,
then we have $(c,\rho)\in b$. But we have $(b,\pi)\in a$ and thus $c\in\mbox{Cl}(a)$ and it follows that
$\|c\neps{\cal V}(a)\|=\Pi$.\\
Therefore, $\eta\force c\neps b$~; by hypothesis on $\xi$, we have $\xi\star\eta\ps\pi\in\bbot$.

\smallskip\noindent
ii)~~Let $a,\vec{c}$ be individuals in ${\cal M}$~; we must show \
$\III\,\force A\eps{\cal Q}(a)$, where $A=\{y\eps a\;;\;F[y,\vec{c}]\}$.\\
We have \ $A=\{(b,\xi\ps\pi)\;;\;(b,\pi)\in a,\;\xi\force F[b,\vec{c}]\}$ and therefore \
$A\subset\mbox{Cl}(a)\fois\Pi$. But we have~:\\
$\|A\neps{\cal Q}(a)\|=\{\pi\in\Pi\;;\;(A,\pi)\in{\cal Q}(a)\}=\Pi$ \ and therefore \
$\III\,\force A\eps{\cal Q}(a)$.

\qed

\smallskip\noindent
We can now define the function symbols $\ov{\bigcup}$ and $\ov{\cal P}$ by setting~:

\smallskip
\centerline{$\ov{\bigcup}x=\{z\eps{\cal V}(x)\;;\;(\ex y\eps x)\,z\eps y\}$~; \
$\ov{\cal P}(x)=\{y\eps{\cal Q}(x)\;;\;y\subseteq x\}$.}

\smallskip\noindent
{\bfseries Collection.}\\
We shall use in the following, function symbols associated with a strong form
of the \emph{collection scheme}.\\
In order to define these function symbols, it is convenient to decompose them, which is done in
theorems~\ref{choix_ensembliste}, \ref{compr_fort} and  \ref{image}.

\begin{theorem}\label{choix_ensembliste}
For each formula $F(x,\vec{z})$ of \ZFe, we have~:\\
\centerline{$\force\pt\vec{z}\left(\ex x\,F(x,\vec{z})\to(\ex x\eps\phi_F(\vec{z}))F(x,\vec{z})\right)$~; \
$\force\pt\vec{z}(\pt x\eps\phi_F(\vec{z}))F(x,\vec{z})$}

\noindent
where $\phi_F$ is a function symbol defined in ${\cal M}.$
\end{theorem}\noindent
We show \ $\lbd x(x)\,\III\force\pt x(x\eps\Phi_F(\vec{z})\to F(x,\vec{z}))\to\pt x\,F(x,\vec{z})$ \
where the function symbol $\Phi_F$ is defined as follows~:

\smallskip\noindent
By means of the collection scheme in ${\cal M}$, we define a function symbol $\Psi(\vec{z})$ such that~:\\
$\|\pt x\,F(x,\vec{z})\|=\bigcup_{x\in\Psi(\vec{z})}\|F(x,\vec{z})\|$ \ and we set \
$\Phi_F(\vec{z})=\Psi(\vec{z})\fois\Pi$.\\
Let $\xi\force\pt x(x\eps\Phi_F(\vec{z})\to F(x,\vec{z}))$ \ and \ $\pi\in\|\pt x\,F(x,\vec{z})\|$.\\
Then \ $\pi\in\|F(x,\vec{z})\|$ for some $x\in\Psi(\vec{z})$, and therefore \ $\III\,\force x\eps\Phi_F(\vec{z})$
and $\xi\star\III\ps\pi\in\bbot$.

\smallskip\noindent
Therefore, by replacing $F$ with $\neg F$, we have \
$\force\ex x\,F(x,\vec{z})\to(\ex x\eps\Phi_{\neg F}(\vec{z}))\,F(x,\vec{z})$.

\smallskip\noindent
Thus, we only need to set \ $\phi_F(\vec{z})=\{x\eps\Phi_{\neg F}(\vec{z})\;;\;F(x,\vec{z})\}$.

\qed

\begin{theorem}\label{compr_fort}
For every formula $F(y,\vec{z})$ of \ZFe, we have~:\\
\centerline{$\force\pt\vec{z}\left(\ex x\pt y(F(y,\vec{z})\to y\eps x)\to
\pt y(F(y,\vec{z})\dbfl y\eps\gamma_F(\vec{z}))\right)$}

\noindent
where $\gamma_F$ is a function symbol defined in ${\cal M}.$
\end{theorem}\noindent
By theorem~\ref{choix_ensembliste}, we have~:

\smallskip\noindent
\centerline{$\force\pt\vec{z}\left(\ex x\pt y(F(y,\vec{z})\to y\eps x)\to
(\ex x\eps\phi(\vec{z}))\pt y(F(y,\vec{z})\to y\eps x)\right)$}

\smallskip\noindent
where $\phi$ is a function symbol. Therefore we have, by definition of \ $\ov{\bigcup}\phi(\vec{z})$~:

\smallskip\noindent
\centerline{$\force\pt\vec{z}\left(\ex x\pt y(F(y,\vec{z})\to y\eps x)\to
\pt y(F(y,\vec{z})\to y\eps\ov{\bigcup}\phi(\vec{z}))\right)$.}

\smallskip\noindent
Now, we only need to set \ $\gamma_F(\vec{z})=\{y\eps\ov{\bigcup}\phi(\vec{z})\;;\;F(y,\vec{z})\}$ \
(comprehension scheme). 

\qed

\smallskip\noindent
When the hypothesis \ $\ex x\pt y(F(y,\vec{z})\to y\eps x)$ is satisfied, we say that \emph{the formula
$F(y,\vec{z})$ defines a set}.\\
For the function symbol $\gamma_F(\vec{z})$, we shall use the more intuitive notation \
$\{y\;;\;F(y,\vec{z})\}$, where $y$ is a bounded variable.

\begin{theorem}\label{image}\ \\
Let $f(x,\vec{z})$ be a $(n+1)$-ary function symbol (defined in ${\cal M}$). Then, we have~:\\
\centerline{$\force\pt a\pt y\pt\vec{z}\left(y\eps\phi_f(a,\vec{z})\dbfl(\ex x\eps a)(y=f(x,\vec{z}))\right)$}

\noindent
where $\phi_f$ is a $(n+1)$-ary function symbol.
\end{theorem}\noindent
We define, in ${\cal M}$, the symbol $\phi_f$ as follows~:\\
Let $a_0,y_0,\vec{z}_0$ be fixed individuals in ${\cal M}$~; we set \
$\phi_f(a_0,\vec{z}_0)=\{(f(x,\vec{z}_0),\pi)\;;\;(x,\pi)\in a_0\}$.\\
Then, we have immediately \
$\|y_0\neps\phi_f(a_0,\vec{z}_0)\|=\|\pt x(y_0=f(x,\vec{z}_0)\hto x\neps a_0)\|$. Therefore~:\\
$\force\pt x(y_0=f(x,\vec{z}_0)\hto x\neps a_0)\dbfl y_0\neps\phi_f(a_0,\vec{z}_0)$ \
which gives the desired result.

\qed

\smallskip\noindent
{\small{\bfseries{}Remark.} The \emph{connective} $\,\hto$ is defined in~\cite{krivine6,krivine7}.}

\smallskip\noindent
For the function symbol $\phi_f(a,\vec{z})$, we shall use the more intuitive notation \
$\{f(x,\vec{z})\;;\;x\eps a\}$, where $x$ is a bounded variable. We call it
\emph{image of $a$ by the function $f(x)$}.

\subsection*{Symbols for characteristic functions}\noindent
Let $R(x_1,\ldots,x_n)$ be a $n$-ary relation defined in ${\cal M}$. Its \emph{ characteristic function},
with values in $\{0,1\}$, will be denoted by $\lbr R(x_1,\ldots,x_n)\rbr$. Therefore, we have~: \
${\cal M}\models\pt\vec{x}(R(\vec{x})\dbfl\lbr R(\vec{x})\rbr=1)$.\\
Therefore, in the realizability model ${\cal N}$, the function symbol $\lbr R(\vec{x})\rbr$ takes its
values in~$\gl2$.

\smallskip\noindent
The theorem~\ref{cons_bf} below, shows that, if a binary relation $y\prec x$ is well founded
in~${\cal M}$, then the relation $\lbr y\prec x\rbr=1$ is  well founded in ${\cal N}$.

\subsection*{Miscellaneous symbols}\noindent
In the following, we shall use some function symbols, the definition and properties of which
are given in~\cite{krivine7}. We simply recall their definition below.

\begin{itemize}
\item The unary function symbol $\gimel$, defined in ${\cal M}$ by $\gimel x=x\fois\Pi$.\\
For any individual $E$ of ${\cal M}$, the \emph{restricted quantifier} $\pt x^{\gl E}$ is defined
in \cite{krivine6} or \cite{krivine7} by~:\\
$\|\pt x^{\gl E}F[x]\|=\bigcup_{x\in E}\|F[x]\|$ \ and we have \
$\force\pt x^{\gl E}F[x]\dbfl\pt x(x\eps\gl E\to F[x])$.\\
In the realizability model ${\cal N}$, the formula $x\eps\gl E$ may be intuitively understood as
\emph{``$x$ is of type $E$''}. For instance, $\gl2$ may be considered as \emph{the type of booleans} and
$\gl\NN$ as \emph{the type of integers}.

\item the function symbols $\et,\ou,\non$, with domains $\{0,1\}\fois\{0,1\}$ and $\{0,1\}$, and values
in $\{0,1\}$, are defined in ${\cal M}$ by means of the usual truth tables.\\
These functions define, in ${\cal N}$, a structure of Boolean algebra on $\gl2$.\\
We call it the \emph{characteristic Boolean algebra} of the realizability model ${\cal N}$.

\item  a binary function symbol with domain $\{0,1\}\fois{\cal M}$, denoted by $(\alpha,x)\mapsto\alpha x$,
by setting~:\\
\hspace*{15em}$0x=\vide$~; $1x=x$.\\
In the model ${\cal N}$, the domain of this function is $\gl2\fois{\cal N}$.

\smallskip
\item a binary function symbol $\,\sqcup$ with domain ${\cal M}\fois{\cal M}$, by setting $x\sqcup y=x\cup y$.\\
{\small{\bfseries{}Remark.}
The extension of this function to the model ${\cal N}$  \emph{is not} the union $\cup$,
which explains the use of another symbol.}
\end{itemize}

\begin{lemma}[Linearity]\label{sqcup}\ \\
Let $f$ be a binary function symbol, defined in ${\cal M}$. Then, we have~:\\
i)~~$\III\,\force\pt\alpha^{\gl2}\pt x\pt y(\alpha f(x,y)=\alpha f(\alpha x,y))$.\\
ii)~~Moreover, if $f(\vide,\vide)=\vide$, then~:\\
$\III\,\force\pt\alpha^{\gl2}\pt\alpha'^{\gl2}
\pt x\pt y\pt x'\pt y'\left(\alpha\et\alpha'=0\hto
f(\alpha x\sqcup\alpha' x',\alpha y\sqcup\alpha' y')=\alpha f(x,y)\sqcup\alpha' f(x',y')\right)$.
\end{lemma}\noindent
{\small{\bfseries{}Remark.} The \emph{connective} $\,\hto$ is defined in~\cite{krivine6,krivine7}.}

\smallskip\noindent
It suffices to check~:\\
for (i) the two cases $\alpha=0,1$~;\\
for (ii) the three cases $(\alpha,\alpha')=(0,0),(0,1),(1,0)$~;\\
which is is trivial.

\qed

\section*{Well founded relations}\noindent
In this section, we study properties of well founded relations in ${\cal N}$. All the results
obtained here are, of course, trivial in ZF\/. The difficulties come from the fact that the relation
$\veps$ of strong membership, does not satisfy extensionality.

\smallskip\noindent
Given a binary relation $\,\prec$, an individual $a$ is said \emph{minimal for $\prec$}, if we have \
$\pt x\,\neg(x\prec a)$.\\
The binary relation $\,\prec$ is called \emph{well founded} if we have~:

\smallskip\noindent
\centerline{$\pt X\left(\pt x(\pt y(y\prec x\to y\neps X)\to x\neps X)\to\pt x(x\neps X)\right)$.}

\smallskip\noindent
The intuitive meaning is that each non empty individual $X$ has an $\veps$-element minimal for~$\prec$.

\begin{theorem}\label{bf_ens}\ \\
If the relation $x\prec y$ is well founded then, for every formula $F[x,\vec{z}]$ of \ZFe, we have~:\\
\centerline{$\pt\vec{z}\left(\pt x(\pt y(y\prec x\to F[y,\vec{z}])\to F[x,\vec{z}])\to
\pt x\,F[x,\vec{z}]\right)$.}
\end{theorem}\noindent
Proof by contradiction~; we consider, in~${\cal N}$, an individual $a$ and a formula $G[x]$
such that~:\\
$(1)$\hspace{10em}$G[a]\;;\;\pt x\left(G[x]\to\ex y\{G[y],y\prec x\}\right)$.\\
We apply the axiom scheme of infinity of \ZFe~:\\
$(2)$\hspace{10em}$\ex c\left\{a\eps c,(\pt x\eps c)\left(\ex y\,F(x,y)\to(\ex y\eps c)F(x,y)\right)\right\}$\\
by setting $F(x,y)\equiv G[x]\land G[y]\land y\prec x$. Let $b=\{x\eps c\;;\;G(x)\}$~;
by $(1)$ and $(2)$, we get~$a\eps b$.\\
We obtain a contradiction with the hypothesis, by showing \ $(\pt x\eps b)(\ex y\eps b)(y\prec x)$.\\
Therefore, we suppose $x\eps c$ and $G[x]$~; by $(2)$, we have~:\\
\centerline{$\ex y\{G[x],G[y],y\prec x\}\to(\ex y\eps c)\{G[x],G[y],y\prec x\}$.}

\noindent
By $G[x]$ and $(1)$, we have $\ex y\{G[x],G[y],y\prec x\}$.\\
Therefore, we have $(\ex y\eps c)\{G[y],y\prec x\}$, hence the result.

\qed

\smallskip\noindent
Therefore, in order to show $\pt x\,F[x]$, it suffices to show
$\pt x\left(\pt y(y\prec x\to F[y])\to F[x]\right)$.\\
Then, we say that we have shown $\pt x\,F[x]$ \emph{by induction on $x$, following
the well founded relation $\,\prec$}.

\begin{theorem}\label{bf-in}
The binary relation $x\in y$ is well founded.
\end{theorem}\noindent
We must show \ $\pt x(\pt y(y\in x\to y\neps X)\to x\neps X)\to\pt x(x\neps X)$.\\
We apply theorem~\ref{bf_ens} to the well founded relation $x\eps y$ and the formula
$F[x]\equiv x\notin X$.\\
This gives~: \ $\pt x(\pt y(y\eps x\to y\notin X)\to x\notin X)\to\pt x(x\notin X)$.\\
Now, we have immediately $\force x\notin X\to x\neps X$. Thus, it remains to show~:\\
$\force\pt x(\pt y(y\in x\to y\neps X)\to x\neps X)\to
\pt x(\pt y(y\eps x\to y\notin X)\to x\notin X)$.\\
But we have $x\notin X\equiv\pt x'(x'\simeq x\to x'\neps X)$. Therefore, we need to show~:\\
$\force\pt x(\pt y(y\in x\to y\neps X)\to x\neps X),
\pt y(y\eps x\to y\notin X),x'\simeq x\to x'\neps X$~; or else~:\\
$\force\pt y(y\eps x\to y\notin X),x'\simeq x\to\pt y(y\in x'\to y\neps X)$.\\
Now, from \ $x'\simeq x,y\in x'$, we deduce $y\in x$. Thus, there is some $y'\simeq y$ such that $y'\eps x$.\\
Then, from $\pt y(y\eps x\to y\notin X)$, we deduce $y'\notin X$, and therefore \ $y\neps X$.

\qed

\smallskip\noindent
For instance, in the following, we shall use the fact that, if there is an ordinal $\rho$ such that $F[\rho]$,
then there exists a least such ordinal, for any formula $F[\rho]$ \emph{written in the language of~~\ZFe}.
This results from theorem~\ref{bf-in}.

\subsubsection*{Preservation of well-foundedness}

\begin{theorem}\label{cons_bf}
Let \ $\prec$ be a well founded binary relation, defined in the ground model ${\cal M}$.
Then, the relation \ $\lbr y\prec x\rbr=1$ is well founded in ${\cal N}$. In fact, we have~:\\
\centerline{$\Y\force\pt X\left(\pt x(\pt y(\lbr y\prec x\rbr=1\hto y\neps X)\to x\neps X)\to
\pt x(x\neps X)\right)$}

\noindent
where \ $\Y=(\lbd x\lbd f(f)(x)xf)\lbd x\lbd f(f)(x)xf$ (Turing fixpoint combinator).
\end{theorem}\noindent
Let $\xi\in\Lbd$ be such that \ $\xi\force\pt x(\pt y(\lbr y\prec x\rbr=1\hto y\neps X_0)\to x\neps X_0)$,
$X_0$ being any individual in~${\cal M}$. We set
$F[x]\equiv(\pt\pi\in\|x\neps X_0\|)(\Y\star\xi\ps\pi\in\bbot)$, and we have to show~$\pt x\,F[x]$.\\
Since \ $\prec$ is a well founded relation, it suffices to show \
$\pt x\left(\pt y(y\prec x\to F[y])\to F[x]\right)$, or else $\neg F[x_0]\to(\ex y\prec x_0)\neg F[y]$,
for any individual $x_0$.\\
By the hypothesis $\neg F[x_0]$, there exists $\pi_0\in\|x_0\neps X_0\|$ such that
$\Y\star\xi\ps\pi_0\notin\bbot$ and therefore, we have $\xi\star\Y\xi\ps\pi_0\notin\bbot$.\\
By hypothesis on $\xi$, we deduce \ $\Y\xi\nforce\pt y(\lbr y\prec x_0\rbr=1\hto y\neps X_0)$.\\
Thus, there exists $y_0\prec x_0$ such that $\Y\xi\nforce y_0\neps X_0$.\\
Therefore, we have
$(\ex\pi\in\|y_0\neps X_0\|)(\Y\star\xi\ps\pi\notin\bbot)$, that is $\neg F[y_0]$.

\qed

\subsection*{Definition of a rank function}\noindent
{\bfseries Definition.} A {\em function with domain $D$} is an individual $\phi$ such that~:\\
$(\pt z\eps\phi)(\ex x\eps D)\ex y(z=(x,y))$~; \ $(\pt x\eps D)\ex y((x,y)\eps\phi)$~;\\
$\pt x\pt y\pt y'((x,y)\eps\phi,(x,y')\eps\phi\to y=y')$.

\smallskip\noindent
Let $\phi$ be a function with domain $D$ and $F[y,\vec{z}]$ a formula of \ZFe. Then, the formula~:\\
$\ex y\{(x,y)\eps\phi,\,F[y,\vec{z}]\}$ \ is denoted by \ $F[\phi(x),\vec{z}]$.

\smallskip\noindent
{\small{\bfseries{}Remark.} Beware, despite the same notation $\phi(x)$, it is not a function symbol.}

\smallskip\noindent
By means of theorem~\ref{compr_fort}, we define the binary function symbol Im by setting~:\\
\centerline{$\mbox{Im}(\phi,D)=\{y\;;\;(\ex x\eps D)\,(x,y)\eps\phi\}$.}

\smallskip\noindent
When $\phi$ is a function with domain $D$, we shall use, for $\mbox{Im}(\phi,D)$, the more intuitive
notation $\{\phi(x)\;;\;x\eps D\}$, which we call \emph{image of the function $\phi$}.

\smallskip\noindent
Let $D'\subseteq D$, that is $\pt x(x\neps D\to x\neps D')$~; \emph{a restriction of $\phi$ to $D'$} 
is, by definition, a function $\phi'$ with domain $D'$ such that \ $\phi'\subseteq\phi$.\\
For instance, $\{z\eps\phi\;;\;(\ex x\eps D')\ex y(z=(x,y))\}$ is a restriction of $\phi$ to~$D'$.\\
If $\phi'_0,\phi'_1$ are both restrictions of $\phi$ to $D'$, then $\phi'_0\cong\phi'_1$.

\smallskip\noindent
{\bfseries Definition.}\\
A binary relation $\,\prec$ is called \emph{ranked}, if we have $\pt x\ex y\pt z(z\prec x\to z\eps y)$,
in other words~: the minorants of any individual form a set.\\
By theorem~\ref{compr_fort}, if the relation $\prec$ is ranked and defined by a formula
$P[x,y,\vec{u}]$ of \ZFe\ with parameters $\vec{u}$ in ${\cal N}$, we have~:\\
${\cal N}\models\pt x\pt y(x\prec y\dbfl x\eps f(y,\vec{u}))$, \emph{for some symbol of function $f$,
defined in ${\cal M}$}.

\smallskip\noindent
In what follows, we suppose that $\,\prec$ is a ranked \emph{transitive} binary relation.

\smallskip\noindent
A function $\phi$ with domain $\{x\;;\;x\prec a\}$ will be called \emph{$a$-inductive for $\,\prec$},
if we have~:\\
$\phi(x)\simeq\{\phi(y)\;;\;y\prec x\}$ for every $x\prec a$. In other words~:\\
$(\pt x\prec a)(\pt y\prec x)\,\phi(y)\in\phi(x)$~;
$(\pt x\prec a)(\pt z\eps\phi(x))(\ex y\prec x)\,z\simeq\phi(y)$.

\smallskip\noindent
If $\phi$ is $a$-inductive for $\,\prec$, we set \ $\OOO(\phi,a)=\{\phi(x)\;;\;x\prec a\}$
(image of $\phi$).

\begin{lemma}\label{2-induct}Let $\phi,\phi'$ be two functions, $a$-inductive for $\,\prec$. Then~:\\
i)~~$\phi(x)\simeq\phi'(x)$ for every $x\prec a$.\\
ii)~~$\OOO(\phi,a)\simeq\OOO(\phi',a)$.\\
iii)~~$(\pt x\prec a)\mbox{On}(\phi(x))$~; \ $\OOO(\phi,a)$ is an ordinal, called \emph{ordinal of $\phi$}.
\end{lemma}\noindent
i)~~Proof by induction on $\phi(x)$, following $\in$~: \ if $u\eps\phi(x)$, then $u\simeq\phi(y)$ with $y\prec x$.\\
Since $\phi(y)\in\phi(x)$, we have \ $\phi(y)\simeq\phi'(y)$ by the induction hypothesis~;\\
therefore $\phi(y)\in\phi'(x)$ and $\phi(x)\subset\phi'(x)$.\\
Conversely, \ if $u\eps\phi'(x)$, then $u\simeq\phi'(y)$ with $y\prec x$. Thus, we have $\phi(y)\in\phi(x)$,
and therefore \ $\phi(y)\simeq\phi'(y)$ by the induction hypothesis~; therefore
$u\in\phi(x)$ and $\phi'(x)\subset\phi(x)$.

\smallskip\noindent
ii)~~Immediate, by (i).

\smallskip\noindent
iii)~~We show $\mbox{On}(\phi(x))$ by induction on $\phi(x)$, for the well founded relation $\in$~:\\
If $u\eps\phi(x)$, we have $u\simeq\phi(y)$ with $y\prec x$~; therefore, we have $\mbox{On}(u)$ by the
induction hypothesis. If $v\eps u$, then $v\eps\phi(y)$, therefore $v\simeq\phi(z)$ with $z\prec y$~;
therefore $v\in\phi(x)$.\\
It follows that $\phi(x)$ is a transitive set of ordinals, thus an ordinal.\\
Then, $\OOO(\phi,a)$ is also a transitive set of ordinals, and therefore an ordinal.

\qed

\begin{lemma}\label{ind_rest}
If $\phi$ is $a$-inductive for $\,\prec$, and if $b\prec a$, then every restriction $\psi$ of $\phi$ to the
domain $\{x\;;\; x\prec b\}$ is a $b$-inductive function for~$\,\prec$.
\end{lemma}\noindent
Indeed, we have,
$\psi(x)=\phi(x)\simeq\{\phi(y)\;;\;y\prec x\}\simeq\{\psi(y)\;;\;y\prec x\}$.

\qed

\smallskip\noindent
By means of theorem~\ref{choix_ensembliste}, we define a unary function symbol $\Phi$, such that~:\\
$\pt x(\pt f\eps\Phi(x))(f\mbox{ is a }x\mbox{-inductive function})$~;\\
$\pt x\pt f\bigg(f\mbox{ is a }x\mbox{-inductive function}\to\ex f(f\eps\Phi(x))\bigg)$.

\smallskip\noindent
In other words, $\Phi(x)$ is a set of $x$-inductive functions, which is non void if there exists
at least one such function.\\
Finally, we define the unary function symbol Rk, using theorem~\ref{image}, by setting~:\\
\centerline{$\mbox{Rk}(x)=\ov{\bigcup}\{\OOO(f,x)\;;\;f\eps\Phi(x)\}$}
\noindent
(the symbol $\ov{\bigcup}$ is defined after theorem~\ref{parties}).\\
Therefore, $\mbox{Rk}(x)$ is the union of the ordinals of the $x$-inductive functions in the set~$\Phi(x)$.\\
Since all these ordinals are extensionally equivalent, by lemma~\ref{2-induct}(ii),
their union $\mbox{Rk}(x)$ is also an equivalent ordinal.

\smallskip\noindent
{\small{\bfseries{}Remarks.}\\ 
If there exists no $x$-inductive function, then $\mbox{Rk}(x)$ is void.\\
The function symbols $\OOO,\Phi,\mbox{Rk}$ have additional arguments, which are the
parameters $\vec{u}$ of the formula $P[x,y,\vec{u}]$ which defines the relation $y\prec x$.}

\smallskip\noindent
We suppose now that $\prec$ is a  ranked transitive relation, which is \emph{well founded}.
It is therefore a \emph{strict ordering}.

\begin{lemma}\label{restr}
Every restriction of Rk to the domain $\{x\;;\;x\prec a\}$ is an $a$-inductive function for \ $\prec$.
\end{lemma}\noindent
Proof by induction on $a$, following $\,\prec$.\\
Let $f$ be a restriction of Rk to the domain $\{x\;;\;x\prec a\}$ and let \ $x\prec a$.
We must show that \ $f(x)\simeq\{f(y)\;;\;y\prec x\}$, in other words, that we have~:\\
\centerline{$\mbox{Rk}(x)\simeq\{\mbox{Rk}(y)\;;\;y\prec x\}$.}

\noindent
Let $\psi$ be any restriction of Rk to the domain $\{y\;;\;y\prec x\}$.
By the induction hypothesis, $\psi$ is a $x$-inductive function for $\,\prec$.\\
We now show that $\mbox{Rk}(x)\simeq\{\mbox{Rk}(y)\;;\;y\prec x\}$~:

\smallskip\noindent
i)~~If $u\eps\mbox{Rk}(x)$ then $u\eps\OOO(\phi,x)$ for some function $\phi$ which is $x$-inductive
for $\prec$, \emph{provided that there exists such a function}. Now, there exists effectively one,
otherwise $\mbox{Rk}(x)$ would be void.\\
Therefore, by definition of $\OOO(\phi,x)$, we have $u=\phi(y)$ with $y\prec x$.
But $\mbox{Rk}(y)\simeq\phi(y)$, since $\phi,\psi$ are both $x$-inductive functions for $\prec$,
and $\psi(y)=\mbox{Rk}(y)$ (lemma~\ref{2-induct}(i)).\\
Therefore, we have $u\simeq\mbox{Rk}(y)$, with $y\prec x$.\\
ii)~~Conversely, if $y\prec x$, then $\mbox{Rk}(y)=\psi(y)$. Let $\phi\eps\Phi(x)$~; then
$\phi,\psi$ are $x$-inductive for~$\prec$~; therefore $\phi(y)\simeq\psi(y)$ (lemma~\ref{2-induct}(i)).\\
Now \ $\phi(y)\eps\OOO(\phi,x)$, and therefore $\phi(y)\eps\mbox{Rk}(x)$ by definition of $\mbox{Rk}(x)$.\\
It follows that \ $\mbox{Rk}(y)=\psi(y)\in\mbox{Rk}(x)$.

\qed

\begin{theorem}\label{D-induct}
We have \ $\mbox{Rk}(x)\simeq\{\mbox{Rk}(y)\;;\;y\prec x\}$ for every $x$.
\end{theorem}\noindent
Proof by induction on $x$, following $\prec$~; let $\psi$ be any restriction of Rk to the domain
$\{y\;;\;y\prec x\}$. By lemma~\ref{restr}, $\psi$ is a $x$-inductive function for $\,\prec$.\\
Then, we finish the proof, by repeating paragraphs (i) and (ii) of the proof of lemma~\ref{restr}.

\qed

\smallskip\noindent
Rk is called the \emph{ rank function} of the ranked, well founded and transitive relation $\,\prec$.\\
$\mbox{Rk}(x)$ is, for every $x$, a representative of the ordinal of any $x$-inductive function for~$\,\prec$.

\smallskip\noindent
The values of the rank function Rk form an initial segment of On, which we shall call
\emph{the image of \ Rk}. \ It is therefore, \emph{either an ordinal, or the whole of On.}

\begin{proposition}\label{prec_01}
Let $\prec_0,\prec_1$ be two ranked transitive well founded relations, and $f$ a function such that \
$\pt x\pt y(x\prec_0 y\to f(x)\prec_1f(y))$.\\
If $\mbox{Rk}_0,\mbox{Rk}_1$ are their rank functions, then we have
$\pt x\left(\mbox{Rk}_0(x)\le\mbox{Rk}_1(f(x))\right)$, and the image of $\mbox{Rk}_0$ is an initial
segment of the image of $\mbox{Rk}_1$.
\end{proposition}\noindent
We show immediately $\pt x\left(\mbox{Rk}_0(x)\le\mbox{Rk}_1(f(x))\right)$ by induction following $\prec_0$.
Hence the result, since the image of a rank function is an initial segment of On.

\qed

\section*{An ultrafilter on $\gl2$}\noindent
In all of the following, we write $y<x$ for $y\in\mbox{Cl}(x)$ in ${\cal M}$, where $\mbox{Cl}(x)$
denotes the transitive closure of $x$. It is a strict well founded ordering (many other such
orderings would do the job, for instance the relation \ rank$(y)<$ rank$(x)$).\\
The binary function symbol $\lbr y<x\rbr$ is therefore defined in ${\cal N}$, with values in $\gl2$.\\
By theorem~\ref{cons_bf}, the binary relation $\lbr y<x\rbr=1$ is well founded in ${\cal N}$.

\begin{theorem}\label{uf_gl2}
$\force\mbox{There exists an ultrafilter }{\cal D}\mbox{ on }\gl2$, which is defined as follows~:\\
${\cal D}=\{\alpha\eps\gl2\;;\mbox{ the relation }\lbr y<x\rbr\ge\alpha\mbox{ is well founded}\;\}$.
\end{theorem}\noindent
{\small{\bfseries{}Remark.} By lemma~\ref{sqcup}, the formula $\lbr y<x\rbr\ge\alpha$
may be written $\lbr\alpha y<\alpha x\rbr=\alpha$.}

\smallskip\noindent
The formula $\alpha\eps{\cal D}$, which we shall also write ${\cal D}[\alpha]$, is therefore~:\nopagebreak

\smallskip\noindent
\centerline{${\cal D}[\alpha]\equiv\pt X\left(\pt x(\pt y(\lbr y<x\rbr\ge\alpha\hto y\neps X)\to x\neps X)
\to\pt x(x\neps X)\right)$}

\smallskip\noindent
{\small{\bfseries{}Remark.} We have~:\\
${\cal D}[1]\equiv\pt X\left(\pt x(\pt y(\lbr y<x\rbr=1\hto y\neps X)\to x\neps X)
\to\pt x(x\neps X)\right)$.\\
${\cal D}[0]\equiv\pt X((\vide\neps X\to \vide\neps X)\to \vide\neps X)$.}

\smallskip\noindent
We have immediately~: \ $\lbd x\,x\;\III\force\neg{\cal D}[0]$~; $\Y\force{\cal D}[1]$~;\\
$\III\,\force\pt\alpha^{\gl2}\pt\beta^{\gl2}\left(\alpha\le\beta\hto
({\cal D}[\alpha]\to{\cal D}[\beta])\right)$
(more precisely~: \ $\|{\cal D}[1]\|\subset\|{\cal D}[0]\|$).

\smallskip\noindent
Therefore, in order to prove theorem~\ref{uf_gl2}, it suffices to show~:\\
$\force\pt\alpha^{\gl2}\pt\beta^{\gl2}\left(\alpha\et\beta=0\hto
({\cal D}[\alpha\ou\beta]\to{\cal D}[\alpha]\lor{\cal D}[\beta])\right)$~;
see theorem~\ref{bf_ou}~;\\
$\force\pt\alpha^{\gl2}\pt\beta^{\gl2}
\left(\alpha\et\beta=0\hto({\cal D}[\alpha],{\cal D}[\beta]\to\bot)\right)$ ~; or even only~:\\
$\force\pt\alpha^{\gl2}({\cal D}[\alpha],{\cal D}[\non\alpha]\to\bot)$~;
see theorem~\ref{bf_et}.

\smallskip\noindent
{\bfseries Notation.} For $\alpha\eps\gl2$, we shall write \ $x<_\alpha y$ \ for \ $\lbr x<y\rbr\ge\alpha$.

\begin{theorem}\label{bf_ou}\ \\
i)~~$\force\pt\alpha^{\gl2}\pt\beta^{\gl2}\left(\alpha\et\beta=0\hto
({\cal D}[\alpha\ou\beta]\to{\cal D}[\alpha]\lor{\cal D}[\beta])\right)$.\\
ii)~~$\force\pt\alpha^{\gl2}\pt\beta^{\gl2}\left(
{\cal D}[\alpha\ou\beta]\to{\cal D}[\alpha]\lor{\cal D}[\beta]\right)$.
\end{theorem}\noindent
i)~~Let $\alpha,\beta\eps\gl2$ be such that $\alpha\et\beta=0,
\neg{\cal D}[\alpha],\neg{\cal D}[\beta]$. We have to show \ $\neg{\cal D}[\alpha\ou\beta]$.\\
By hypothesis on $\alpha$ and $\beta$, there exists individuals $a_0,A$ (resp. $b_0,B$) such that $a_0\eps A$ (resp. $b_0\eps B$) and $A$ (resp. $B$) has no minimal $\veps$-element for~$<_\alpha$ (resp. for $<_\beta$).
We set~:\\
\centerline{$c_0=\alpha a_0\sqcup\beta b_0$ \ and $C=\{\alpha x\sqcup\beta y\;;\;x\eps A,y\eps B\}$.}

\smallskip\noindent
Therefore, we have $c_0\eps C$~; it suffices to show that $C$ has no minimal $\veps$-element for
$<_{\alpha\ou\beta}$. Let $c\eps C,c=\alpha a\sqcup\beta b$, with $a\eps A,b\eps B$. By hypothesis on
$A,B$, there exists $a'\eps A$ and $b'\eps B$ such that $a'<_\alpha a$, $b'<_\beta b$. If we set
$c'=\alpha a'\sqcup\beta b'$, we have $c'\eps C$, as needed. We also have~:\\
$\lbr c'=a'\rbr\ge\alpha,\lbr a'<a\rbr\ge\alpha,\lbr c=a\rbr\ge\alpha$~;
it follows that $\lbr c'<c\rbr\ge\alpha$.\\
In the same way, we have $\lbr c'<c\rbr\ge\beta$ and therefore, finally, $\lbr c'<c\rbr\ge\alpha\ou\beta$.

\smallskip\noindent
ii)~~We set $\beta'=\beta\et(\non\alpha)$~; we have $\alpha\et\beta'=0$ and
$\alpha\ou\beta'=\alpha\ou\beta$. Therefore, we have~:\\
${\cal D}[\alpha\ou\beta]\to{\cal D}[\alpha]\lor{\cal D}[\beta']$.
Now, we have \ $\beta'\le\beta$ and therefore ${\cal D}[\beta']\to{\cal D}[\beta]$.

\qed

\begin{lemma}\label{ppt-eps}\ \\
i)~~$\III\,\force\pt x\pt y(\lbr x<y\rbr\ne1\to x\neps y)$.\\
ii)~~If ${\cal M}\models u\in v$, then \ $\III\,\force u\eps\gl v$.\\
iii)~~$\III\,\force\pt x\pt y\pt\alpha^{\gl2}
\left(\lbr x<y\rbr\ge\alpha\hto\alpha x\eps\gl\mbox{Cl}(\{y\})\right)$.\\
iv)~~$\force\pt x\pt y\left(\lbr x<y\rbr=1\dbfl x\eps\gl\mbox{Cl}(y)\right)$.
\end{lemma}\noindent
Let $a,b$ be two individuals.\\
i)~~Let $\xi\force\lbr a<b\rbr\ne1$, $\pi\in\|a\neps b\|$~;
then $(a,\pi)\in b$, therefore $\lbr a<b\rbr=1$ and $\xi\force\bot$~;\\
therefore $\xi\star\pi\in\bbot$.

\smallskip\noindent
ii)~~Indeed, we have $\|u\neps\gl v\|=\{\pi\in\Pi\;;\;(u,\pi)\in v\fois\Pi\}=\Pi$.

\smallskip\noindent
iii)~~Let $\alpha\in\{0,1\}$ and $a,b\in{\cal M}$ such that $\lbr a<b\rbr\ge\alpha$.\\
If $\alpha=0$, we must show \ $\III\,\force\vide\eps\gl\mbox{Cl}(\{y\})$ which follows from (ii).\\
If $\alpha=1$, then $\lbr a<b\rbr=1$, that is $a\in\mbox{Cl}(b)$, therefore $a\in\mbox{Cl}(\{b\})$.\\
From (ii), it follows that \ $\III\,\force a\eps\gl\mbox{Cl}(\{b\})$.

\smallskip\noindent
iv)~~Indeed, if $a,b$ are individuals of ${\cal M}$, we have trivially~:\\
$\|\lbr a<b\rbr\ne1\|=\|a\neps\gl\mbox{Cl}(b)\|$.

\qed

\begin{lemma}\label{rang1}
The well founded relation $\lbr x<y\rbr=1$ is ranked, and its rank function $\,\RRR$ has for image
the whole of On.
\end{lemma}\noindent
Lemma~\ref{ppt-eps}(iv) shows that this relation is ranked.\\
Let $\rho$ be an ordinal and $r$ an individual $\simeq\rho$. We show, by induction on~$\rho$, that
$\RRR(r)\ge\rho$.\\
Indeed, for every $\rho'\in\rho$, there exists $r'\eps r$ such that $r'\simeq\rho'$.
We have $\,\RRR(r')\ge\rho'$ by induction hypothesis, and $\lbr r'<r\rbr=1$ from lemma~\ref{ppt-eps}(i).
Therefore, we have $\rho'\in\RRR(r)$ by definition of~$\RRR$, and finally $\RRR(r)\ge\rho$.
This shows that the image of $\RRR$ is not bounded in On. Since it is an initial segment, it is the whole of On.

\qed

\begin{theorem}\label{choix_faible}
Let $F(x,y)$ be a formula of \ZFe, with parameters. Then, we have~:\\
$\III\,\force\pt x\pt y\left(\pt\varpi^{\gl\Pi}F(x,f(x,\varpi))\to F(x,y)\right)$\\
for some function symbol $f$, defined dans~${\cal M}$, with domain ${\cal M}\fois\Pi$.
\end{theorem}\noindent
Since the ground model ${\cal M}$ satisfies V = L (or only the \emph{choice principle}), we can define,
in ${\cal M}$, a function symbol $f$ such that~:\\
\centerline{$\pt x\pt y(\pt \varpi\in\Pi)\left(\varpi\in\|F(x,y)\|\to\varpi\in\|F(x,f(x,\varpi))\|\right)$.}

\noindent
Let $a,b$ be individuals, $\xi\force\pt\varpi^{\gl\Pi}F(a,f(a,\varpi))$ and $\pi\in\|F(a,b)\|$.\\ 
Thus, we have \ $\pi\in\|F(a,f(a,\pi))\|$, and therefore $\xi\star\pi\in\bbot$.

\qed

\smallskip\noindent
{\bfseries Definitions.} Let $a$ be any individual of ${\cal N}$ and $\,\kappa$ an ordinal
(therefore, $\kappa$ is not an individual of~${\cal N}$, but an equivalence class for $\simeq$).\\
A \emph{function} or \emph{application} from $\kappa$ into $a$ is, by definition,
a binary relation $R(\rho,x)$ such that~: \
$\pt x\pt x'(\pt \rho,\rho'\in\kappa)\left(R(\rho,x),R(\rho',x'),\rho\simeq\rho'\to x=x')\right)$~;
$(\pt \rho\in\kappa)(\ex x\eps a)R(\rho,x)$.\\
It is an \emph{injection} if we have  \
$\pt x(\pt \rho,\rho'\in\kappa)\left(R(\rho,x),R(\rho',x)\to\rho\simeq\rho'\right)$.\\
A \emph{surjection} from $a$ onto $\kappa$ is a function $f$ of domain $a$ such that~:\\
$(\pt\rho\in\kappa)(\ex x\eps a)\,f(x)\simeq\rho$.

\begin{theorem}\label{surj_ord}\ \\
For any individual $a$, there exists an ordinal $\,\kappa$, such that there is no surjection from $a$
onto $\kappa$.
\end{theorem}\noindent
Let $f$ be a surjection from $a$ onto an ordinal $\rho$. We define a strict ordering relation \
$\prec_f$ \ by setting \ $x\prec_fy$ $\Dbfl$ $x\eps a\land y\eps a\land f(x)<f(y)$. It is clear that
this relation is well founded, that $f$ is an $a$-inductive function, and that $\OOO(f,a)\simeq\rho$.\\
We may consider this relation as a subset of $a\fois a$.\\
By means of the axioms of union, power set and collection given above (theo\-rems~\ref{parties}
à~\ref{image}), we define an ordinal $\kappa_0$, which is the union of the
$\OOO(f,a)$ for all the functions $f$ which are $a$-inductive for some well founded strict ordering
relation on~$a$.\\
In fact, we consider the set~:\\
\centerline{${\cal B}(a)=\{X\eps\ov{\cal P}(a\fois a)\;;\;X\mbox{ is a well founded strict ordering
relation on }a\}$.}\\
Then, we set $\kappa_0=\ov{\bigcup}\{\OOO(f,a)\;;\;X\eps{\cal B}(a),f\eps\Phi(X,a)\}$.\\
In this definition, we use the function symbol $\Phi$, defined after lemma~\ref{ind_rest},
which associates with each  well founded strict ordering relation $X$ on $a$,
a \emph{non void} set of $a$-inductive functions for this relation.

\smallskip\noindent
Then, there exists no surjection from $a$ onto $\kappa_0+1$.

\qed

\smallskip\noindent
We denote by $\Delta$ the first ordinal of ${\cal N}$ such that there is no surjection from
$\gl\Pi$ onto~$\Delta$~:
for every function $\phi$, there exists $\delta\in\Delta$ such that $\pt x^{\gl\Pi}(\phi(x)\not\simeq\delta)$.

\smallskip\noindent
For each $\alpha\eps\gl2$, we denote by ${\cal N}_\alpha$ the class defined by the formula \ $x=\alpha x$.

\begin{lemma}\label{R01_alpha01}
Let  $\alpha_0,\alpha_1\eps\gl2$, $\alpha_0\et\alpha_1=0$ and $R_0$ (resp. $R_1$) be a
functional relation of domain ${\cal N}_{\alpha_0}$ (resp. ${\cal N}_{\alpha_1}$) with values in On.
Then, either $R_0$, or $R_1$, is not surjective onto~$\Delta$.
\end{lemma}\noindent
Proof by contradiction~: we suppose that $R_0$ and $R_1$ are both surjective onto $\Delta$.\\
We apply theorem~\ref{choix_faible} to the formula
$F(x_0,x_1)\equiv\neg(R_0(\alpha_0x_0)\simeq R_1(\alpha_1x_1))$, and we get~:

\smallskip\noindent
\centerline{$\pt x_0\left(\ex x_1(R_0(\alpha_0x_0)\simeq R_1(\alpha_1x_1))
\to\ex\varpi^{\gl\Pi}(R_0(\alpha_0x_0)\simeq R_1(\alpha_1f(x_0,\varpi)))\right)$}

\noindent
where $f$ is a suitable function symbol (therefore \emph{defined in ${\cal M}$}).\\
Replacing $x_0$ with $\alpha_0x_0$, we obtain~:

\centerline{$\pt x_0\left(\ex x_1(R_0(\alpha_0x_0)\simeq R_1(\alpha_1x_1))
\to\ex\varpi^{\gl\Pi}(R_0(\alpha_0x_0)\simeq R_1(\alpha_1f(\alpha_0x_0,\varpi)))\right)$.}

\noindent
But, by lemma~\ref{sqcup}(i), we have \
$\alpha_1f(\alpha_0x,\varpi)=\alpha_1f(\alpha_1\alpha_0x,\varpi)
=\alpha_1f(\vide,\varpi)$.
It follows that~:\\
\centerline{$\pt x_0\left(\ex x_1(R_0(\alpha_0x_0)\simeq R_1(\alpha_1x_1))
\to\ex\varpi^{\gl\Pi}(R_0(\alpha_0x_0)\simeq R_1(\alpha_1f(\vide,\varpi)))\right)$.}

\smallskip\noindent
By hypothesis, we have \ $(\pt\rho\in\Delta)\ex x_0\ex x_1
(\rho\simeq R_0(\alpha_0x_0)\simeq R_1(\alpha_1x_1))$. It follows that~:\\
$(\pt\rho\in\Delta)\ex x_0\ex\varpi^{\gl\Pi}\left(\rho\simeq R_0(\alpha_0x_0)
\simeq R_1(\alpha_1f(\vide,\varpi))\right)$~;  therefore, we have~:\\
$(\pt\rho\in\Delta)\ex\varpi^{\gl\Pi}\left(\rho\simeq R_1(\alpha_1f(\vide,\varpi))\right)$.

\smallskip\noindent
Therefore, the function $\varpi\mapsto R_1(\alpha_1f(\vide,\varpi))$ is a surjection from $\gl\Pi$
onto $\Delta$. But this is a contradiction with the definition de~$\Delta$.

\smallskip\noindent
{\small{\bfseries{}Remark.} We should write $f(\alpha_0,\alpha_1,x_0,\varpi)$ instead of
$f(x_0,\varpi)$, since the function  symbol $f$ depends on the four variables $\alpha_0,\alpha_1,x_0,\varpi$.
In fact, it depends also on the parameters which appear in $R_0,R_1$.
The proof does not change.}

\qed

\begin{corollary}\label{bf_01}
Let $\alpha_0,\alpha_1\eps\gl2$, $\alpha_0\et\alpha_1=0$, and $\prec_0,\prec_1$ be two well
founded ranked strict ordering relations with respective domains ${\cal N}_{\alpha_0},{\cal N}_{\alpha_1}$.
Let $\mbox{Rk}_0$, $\mbox{Rk}_1$ be their rank functions. Then, either the image of $\mbox{Rk}_0$, or
that of $\mbox{Rk}_1$ is an ordinal $\,<\Delta$.
\end{corollary}\noindent
In order to be able to define the rank functions $\mbox{Rk}_0$, $\mbox{Rk}_1$, we consider the relations
$\prec'_0,\prec'_1$, with domain the whole of ${\cal N}$, defined by \
$x\prec'_iy\equiv(x=\alpha_ix)\land(y=\alpha_iy)\land(x\prec_iy)$ for $i=0,1$.\\
These strict ordering relations are well founded and ranked.\\
Their rank functions $\,\mbox{Rk}'_0$, $\mbox{Rk}'_1$ take the value $0$ outside
${\cal N}_{\alpha_0},{\cal N}_{\alpha_1}$ respectively~: indeed, all the individuals outside
${\cal N}_{\alpha_i}$ are minimal for $\,\prec'_i$.

\smallskip\noindent
By lemma~\ref{R01_alpha01}, one of them, $\,\mbox{Rk}'_0$ for instance, is not surjective onto $\Delta$.\\
Since the image of any rank function is an initial segment of On, the image of $\mbox{Rk}_0$ is
an ordinal~$\,<\Delta$.

\qed

\begin{theorem}\label{bf_et}\ \\
i)~~$\force\pt\alpha_0^{\gl2}\pt\alpha_1^{\gl2}
\left(\alpha_0\et\alpha_1=0\hto({\cal D}[\alpha_0],{\cal D}[\alpha_1]\to\bot)\right)$.

\smallskip\noindent
ii)~~$\force\pt\alpha_0^{\gl2}\pt\alpha_1^{\gl2}
\left({\cal D}[\alpha_0],{\cal D}[\alpha_1]\to{\cal D}[\alpha_0\et\alpha_1]\right)$.
\end{theorem}\noindent
i)~~In ${\cal N}$, let $\alpha_0,\alpha_1\eps\gl2$ be such that $\alpha_0\et\alpha_1=0$ and the relations
$\lbr x< y\rbr\ge\alpha_0$, $\lbr x< y\rbr\ge\alpha_1$ be well founded. Therefore, we have $\alpha_0,\alpha_1\ne0,1$.\\
Therefore, the relations $x\prec_iy\equiv(x=\alpha_ix)\land(y=\alpha_iy)\land(\lbr x<y\rbr=\alpha_i)$ for $i=0,1$, are well founded strict orderings.\\
From lemma~\ref{ppt-eps}(iii), it follows that these relations are \emph{ranked}.\\
Now, by lemma~\ref{sqcup}, we have~:
$\force\pt x\pt y\pt\alpha^{\gl2}(\lbr x<y\rbr=1\to\lbr\alpha x<\alpha y\rbr=\alpha)$.\\
But, by lemma~\ref{rang1}, the rank function of the well founded relation $\lbr x<y\rbr=1$
has for image the whole of On. Therefore, by proposition~\ref{prec_01}, the same is true for
the rank functions of the well founded strict order relations $x\prec_0y$ and $x\prec_1y$.\\
But this contradicts corollary~\ref{bf_01}.

\smallskip\noindent
ii)~~We have $\alpha_0\le(\alpha_0\et\alpha_1)\ou(\non\alpha_1)$. Therefore, by ${\cal D}[\alpha_0]$
and theorem~\ref{bf_ou}, we have ${\cal D}[\alpha_0\et\alpha_1]$ or ${\cal D}[\non\alpha_1]$.
But ${\cal D}[\non\alpha_1]$ is impossible, by ${\cal D}[\alpha_1]$ and (i).

\qed

\begin{corollary}\label{surj_Delta}
${\cal D}[\alpha]$ is equivalent with each one of the following propositions~:\\
i)~~There exists a well founded ranked strict ordering relation $\,\prec$ with domain ${\cal N}_\alpha$,
the rank function of which has an image $\,\ge\Delta$.\\
ii)~~There exists a function with domain ${\cal N}_\alpha$ which is surjective onto $\,\Delta$.
\end{corollary}\noindent
${\cal D}[\alpha]\Fl$ (i)~:\\
By definition of ${\cal D}[\alpha]$, the binary relation
$(x=\alpha x)\land(y=\alpha y)\land(\lbr x<y\rbr=\alpha)$ is well founded. By lemma~\ref{ppt-eps}(iii),
this relation is ranked. We have seen, in the proof of theorem~\ref{bf_et}, that the image of its rank
function is the whole of On.

\smallskip\noindent
(i) $\Fl$ (ii)~: obvious.

\smallskip\noindent
(ii) $\Fl{\cal D}[\alpha]$~:\\
Since ${\cal D}$ is an ultrafilter, it suffices to show $\neg{\cal D}[\non\alpha]$.
But, (ii) and ${\cal D}[\non\alpha]$ contradict lemma~\ref{R01_alpha01}.

\qed

\begin{theorem}\label{no_eps_ord}\ \\
If \ $\gl2$ is non trivial, there exists no set, which is totally ordered by $\veps$, the ordinal
of which is~$\ge\Delta$.
\end{theorem}\noindent
Let $\alpha\eps\gl2,\alpha\ne0,1$ and $X$ be a set which is totally ordered by $\eps$, and
equipotent with $\Delta$.\\
Then, we show that the application $x\mapsto\alpha x$ is an injection from $X$ into~${\cal N}_\alpha$~:\\
Indeed, by lemma~\ref{ppt-eps}(i), we have $x\eps y\to\lbr x<y\rbr=1$ and, by lemma~\ref{sqcup}, we have~:\\
$\lbr x<y\rbr=1\to\lbr\alpha x<\alpha y\rbr=\alpha$. Therefore, if $x,y\eps X$ and $x\ne y$, we have,
for instance $x\eps y$, therefore $\lbr\alpha x<\alpha y\rbr=\alpha$ and therefore $\alpha x\ne\alpha y$
since $\alpha\ne0$.

\smallskip\noindent
Thus, there exists a function with domain ${\cal N}_\alpha$ which is surjective onto $\Delta$.
The same reasoning, applied to $\non\alpha$ gives the same result for $\non\alpha$. But this contradicts
lemma~\ref{R01_alpha01}.

\qed

\smallskip\noindent
{\small{\bfseries Remark.} Theorem~\ref{no_eps_ord} shows that it is impossible to define Von Neumann
ordinals in ${\cal N}$, with $\varepsilon$ instead of~$\in$, unless $\gl2$ is trivial, i.e. the realizability
model is, in fact, a forcing model.}

\section*{The model ${\cal M}_{\cal D}$}\noindent
For each formula $F[x_1,\ldots,x_n]$ of ZF\/, we have defined, in the ground model
${\cal M}$, an $n$-ary function symbol with values in $\{0,1\}$, denoted \ by
$\lbr F[x_1,\ldots,x_n]\rbr$, by setting, for any\, individuals $a_1,\ldots,a_n$
of ${\cal M}$~: \ $\lbr F[a_1,\ldots,a_n]\rbr=1$ \ $\Dbfl$ \
${\cal M}\models F[a_1,\ldots,a_n]$.\\
In ${\cal N}$, the function symbol $\lbr F[x_1,\ldots,x_n]\rbr$ takes its values in the Boolean
algebra~$\gl2$.

\smallskip\noindent
We define, in ${\cal N}$, two binary relations $\in_{\cal D}$ and $=_{\cal D}$, by setting~:\\
\centerline{$(x\in_{\cal D}y)\equiv{\cal D}[\lbr x\in y\rbr]$~; \ $(x=_{\cal D}y)\equiv{\cal D}[\lbr x=y\rbr]$.}

\noindent
The class ${\cal N}$, equipped with these relations, will be denoted ${\cal M}_{\cal D}$.

\smallskip\noindent
For each formula $F[\vec{x},y]$ of ZF\/, with $n+1$ free variables $x_1,\ldots,x_n,y$, we can define,
by means of the \emph{choice principle} in ${\cal M}$, an $n$-ary function symbol $f_F$, such that~:\\
\centerline{${\cal M}\models\pt\vec{x}\left(F[\vec{x},f_F(\vec{x})]\to\pt y\,F[\vec{x},y]\right)$~;}

\noindent
$f_F$ is called the \emph{Skolem function} of the formula $F[\vec{x},y]$.

\begin{lemma}\label{F(f_F)}\ \\
i)~~$\III\,\force\pt\vec{x}\pt y\left(\lbr\pt y\,F[\vec{x},y]\rbr\le\lbr F[\vec{x},y]\rbr\right)$\\
ii)~~$\III\,\force\pt\vec{x}\pt y\left(\lbr\pt y\,F[\vec{x},y]\rbr
=\lbr F[\vec{x},f_F(\vec{x})]\rbr\right)$.
\end{lemma}\noindent
Trivial.

\qed

\smallskip\noindent
For each formula $F[\vec{x}]$ of ZF\/, we define, by recurrence on $F$, \emph{a formula
of \ZFe}, which has the same free variables, and that we denote \
${\cal M}_{\cal D}\models F[\vec{x}]$.

\smallskip\noindent
$\bullet$~~$F$ is atomic~:\\
$({\cal M}_{\cal D}\models x_1\in x_2)$ \ is \ $x_1\in_{\cal D}x_2$~; \
$({\cal M}_{\cal D}\models x_1=x_2)$ \ is \ $x_1=_{\cal D}x_2$~; \
$({\cal M}_{\cal D}\models\bot)$ \ is \ $\bot$.

\smallskip\noindent
$\bullet$~~$F\equiv F_0\to F_1$~: then \ $({\cal M}_{\cal D}\models F)$ \ is the formula \
$({\cal M}_{\cal D}\models F_0)\to({\cal M}_{\cal D}\models F_1)$.

\smallskip\noindent
$\bullet$~~$F[\vec{x}]\equiv\pt y\,G[\vec{x},y]$~: then \ $({\cal M}_{\cal D}\models F[\vec{x}])$ \
is the formula \ $\pt y({\cal M}_{\cal D}\models G[\vec{x},y])$.

\begin{lemma}\label{MDF}
For each formula $F[\vec{x}]$ of ZF\/, we have \
$\force\pt\vec{x}\Big(({\cal M}_{\cal D}\models F[\vec{x}])\dbfl
{\cal D}\lbr F[\vec{x}]\rbr\Big)$.
\end{lemma}\noindent
Proof by recurrence on the length of $F$.

\smallskip\noindent
If $F$ is atomic, we have \ $\III\,\force\pt\vec{x}
\Big(({\cal M}_{\cal D}\models F[\vec{x}])\to{\cal D}\lbr F[\vec{x}]\rbr\Big)$\\
and \ $\III\,\force\pt\vec{x}
\Big({\cal D}\lbr F[\vec{x}]\rbr\to({\cal M}_{\cal D}\models F[\vec{x}])\Big)$ \
since \ $({\cal M}_{\cal D}\models F[\vec{x}])$ is identical with ${\cal D}\lbr F[\vec{x}]\rbr$.

\smallskip\noindent
If $F\equiv F_0\to F_1$, the formula $({\cal M}_{\cal D}\models F)\dbfl{\cal D}\lbr F\rbr$ is~:\\
$(({\cal M}_{\cal D}\models F_0)\to({\cal M}_{\cal D}\models F_1))\dbfl{\cal D}\lbr F_0\to F_1\rbr$.\\
Since ${\cal D}$ is an ultrafilter, this formula is equivalent with~:\\
$(({\cal M}_{\cal D}\models F_0)\to({\cal M}_{\cal D}\models F_1))\dbfl
({\cal D}\lbr F_0\rbr\to{\cal D}\lbr F_1\rbr)$, which is a logical consequence of~:\\
$({\cal M}_{\cal D}\models F_0)\dbfl{\cal D}\lbr F_0\rbr$ \ and \
$({\cal M}_{\cal D}\models F_1)\dbfl{\cal D}\lbr F_1\rbr$.\\
Hence the result, by the recurrence hypothesis.

\smallskip\noindent
If $F[\vec{x}]\equiv\pt y\,G[\vec{x},y]$, let $f_G(\vec{x})$ be the Skolem function of $G$.\\
Then, we have \ $({\cal M}_{\cal D}\models\pt y\,G[\vec{x},y])\equiv\pt y({\cal M}_{\cal D}\models G[\vec{x},y])$, and therefore~:\\
$\III\,\force({\cal M}_{\cal D}\models\pt y\,G[\vec{x},y])\to({\cal M}_{\cal D}\models G[\vec{x},f_G(\vec{x})])$.\\
Therefore, by the recurrence hypothesis, we have~:\\
$\force({\cal M}_{\cal D}\models\pt y\,G[\vec{x},y])\to{\cal D}\lbr G[\vec{x},f_G(\vec{x})]\rbr$.\\
Applying lemma~\ref{F(f_F)}(ii), we obtain \
$\force({\cal M}_{\cal D}\models\pt y\,G[\vec{x},y])\to
{\cal D}\lbr\pt y\,G[\vec{x},y]\rbr$.\\
Conversely, by lemma~\ref{F(f_F)}(i), we have
$\force\pt y\left({\cal D}\lbr\pt y\,G[\vec{x},y]\rbr\to{\cal D}\lbr G[\vec{x},y]\rbr\right)$.\\
Therefore, applying the recurrence hypothesis, we obtain~:\\
$\force{\cal D}\lbr\pt y\,G[\vec{x},y]\rbr\to\pt y({\cal M}_{\cal D}\models G[\vec{x},y])$,
and thus, by definition of $({\cal M}_{\cal D}\models \pt y\,G[\vec{x},y])$~:\\
$\force{\cal D}\lbr\pt y\,G[\vec{x},y]\rbr\to({\cal M}_{\cal D}\models \pt y\,G[\vec{x},y])$.

\qed

\begin{theorem}\label{MD_ext_elem}
${\cal M}_{\cal D}$ is an elementary extension of the ground model ${\cal M}$.
\end{theorem}\noindent
Let $F[\vec{a}]$ be a closed formula of ZF\/, with parameters $a_1,\ldots,a_n$ in ${\cal M}$.\\
If ${\cal M}\models F[\vec{a}]$, we have $\lbr F[\vec{a}]\rbr=1$ (by definition),
and therefore, of course, \ $\force{\cal D}\lbr F[\vec{a}]\rbr$.\\
Therefore, by lemma~\ref{MDF}, we have $\force({\cal M}_{\cal D}\models F[\vec{a}])$.\\
If ${\cal M}\nmodels F[\vec{a}]$, then ${\cal M}\models\neg F[\vec{a}]$~;
therefore, we have $\force({\cal M}_{\cal D}\models\neg F[\vec{a}])$.

\qed

\smallskip\noindent
{\small{\bfseries{}Remark.}
Theorem~\ref{MD_ext_elem} is, in fact, true for any ultrafilter on $\gl2$, with the same proof.}

\begin{theorem}\label{sbs_D}
Let \ $\sbs$ be a well founded binary relation, defined in the ground model~${\cal M}$. Then the relation
${\cal D}\lbr x\sbs y\rbr$ is well founded in the realizability model ${\cal N}$.
\end{theorem}\noindent
{\small{\bfseries{}Remark.} Theorem~\ref{sbs_D} is an improvement on theorem~\ref{cons_bf}.}

\smallskip\noindent
{\bfseries Notations.} We shall write \ $x\sbs_{\cal D}y$ \ for \ $\lbr x\sbs y\rbr\eps{\cal D}$.\\
Recall that $x<y$ means $x\in\mbox{Cl}(y)$~; and that \ $x<_\alpha y$ means $\lbr x<y\rbr\ge\alpha$,
for $\alpha\eps\gl2$.

\smallskip\noindent
We define, in the model ${\cal M}$, a binary relation $\ssbs$ on the class $\{0,1\}\fois{\cal M}$,
by setting, for any $\alpha,\alpha'\in\{0,1\}$ and $a,a'$ in ${\cal M}$~:\\
\centerline{$(\alpha',a')\ssbs(\alpha,a)$ $\Dbfl$
$(\alpha'<\alpha)\lor(\alpha=\alpha'=0\land a'< a)\lor(\alpha=\alpha'=1\land a'\sbs a)$.}

\noindent
The relation $\ssbs$ is the \emph{ordered direct sum} of the relations \ $\sbs,<$.\\
It is easily shown that it is \emph{well founded in ${\cal M}$.}

\smallskip\noindent
The binary function symbol associated with this relation, of domain $\{0,1\}\fois{\cal M}$ and values in $\{0,1\}$,
is given by~:\\
\centerline{$\lbr(\alpha',a')\ssbs(\alpha,a)\rbr=(\non\alpha'\et\alpha)
\ou(\non\alpha'\et\non\alpha\et\lbr a'<a\rbr)\ou(\alpha'\et\alpha\et\lbr a'\sbs a\rbr)$.}

\smallskip\noindent
This definition gives, in ${\cal N}$, a binary function symbol with arguments in $\gl2\fois{\cal N}$, and
values in $\gl2$.\\
By theorem~\ref{cons_bf}, \emph{the binary relation $\lbr(\alpha',a')\ssbs(\alpha,a)\rbr=1$ is well
founded in ${\cal N}$.}

\smallskip\noindent
{\em Proof of theorem~\ref{sbs_D}}.\\
Proof by contradiction~: we assume that the binary  relation \ $\sbs_{\cal D}$ \ is not well founded.\\
Thus, there exists $a_0,A_0$ such  that $a_0\eps A_0$ and $A_0$ has no minimal $\veps$-element for \
$\sbs_{\cal D}$.\\
We define, in ${\cal N}$, the class ${\cal X}$ of ordered pairs $(\alpha,x)$, such that~:\\
{\em There exists $X$ such that $x\eps X$ and $X$ has no minimal $\veps$-element, neither for
$\sbs_{\cal D}$ nor for~$<_{\non\alpha}$.}\\
Therefore, the formula ${\cal X}(\alpha,x)$ is~:

\smallskip
\centerline{$\alpha\eps\gl2\land\ex X\bigg\{x\eps X,(\pt u\eps X)\{(\ex v\eps X)(v\sbs_{\cal D}u),
(\ex w\eps X)(w<_{\non\alpha} u)\}\bigg\}$.}

\smallskip\noindent
If $(\alpha,x)$ is in ${\cal X}$, then we have ${\cal D}(\alpha)$~: indeed, the set $X$ is non void
and has no minimal $\veps$-element for $<_{\non\alpha}$. Therefore, we have $\neg{\cal D}(\non\alpha)$,
and thus ${\cal D}(\alpha)$, since ${\cal D}$ is an ultrafilter.

\smallskip\noindent
We obtain the desired contradiction by showing that the class ${\cal X}$ is non void and has no
minimal element for the binary relation \ $\lbr(\alpha',x')\ssbs(\alpha,x)\rbr=1$.

\smallskip\noindent
The ordered pair $(1,a_0)$ is in ${\cal X}$~: indeed, we have $x<_0x$ for every $x$, and therefore $A_0$
has no minimal $\veps$-element for $<_0$.

\smallskip\noindent
Now let $(\alpha,a)$ be in ${\cal X}$~; we search for $(\alpha',a')$ in ${\cal X}$ such that \
$\lbr(\alpha',a')\ssbs(\alpha,a)\rbr=1$.

\smallskip\noindent
By hypothesis on $(\alpha,a)$, there exists $A$ such that $a\eps A$ and $A$ has no minimal $\veps$-element,
neither for \ $\sbs_{\cal D}$ nor for $<_{\non\alpha}$.
Thus, there exists $a^0,a^1\eps A$ such that we have $\,{\cal D}\lbr a^0\sbs a\rbr$ and $a^1<_{\non\alpha}a$.\\
We set $\alpha'=(\alpha\et\lbr a^0\sbs a\rbr)$ and therefore, we have ${\cal D}(\alpha')$.
We set $\beta=\non\alpha'\et\alpha$~; therefore $\alpha',\neg\alpha,\beta$ form a partition of $1$
in the Boolean algebra $\gl2$.\\
We have $\neg\,{\cal D}(\beta)$~; therefore, by definition of ${\cal D}$, the relation $<_\beta$
is not well founded. Thus, there exists $b,B$ such that $b\eps B$ and $B$ has no minimal $\veps$-element
for $<_\beta$. Then, we set~:

\smallskip\noindent
$a'=\alpha'a^0\sqcup(\neg\alpha)a^1\sqcup\beta b$ \ and \
$A'=\{\alpha'x\sqcup(\non\alpha)y\sqcup\beta z~;x,y\eps A,z\eps B\}$.

\smallskip\noindent
Therefore, we have $a'\eps A'$, as needed~; moreover~:\\
$\non\alpha'\et\non\alpha\et\lbr a'<a\rbr=\non\alpha$, since $\non\alpha'\ge\non\alpha$ and
$\lbr a'<a\rbr\ge\non\alpha\et\lbr a^1<a\rbr=\non\alpha$~;\\
$\alpha'\et\alpha\et\lbr a'\sbs a\rbr=\alpha'\et\lbr a'\sbs a\rbr=\alpha'\et\lbr a^0\sbs a\rbr=\alpha'$.\\
By definition of $\lbr(\alpha',a')\ssbs(\alpha,a)\rbr$, it follows that
$\lbr(\alpha',a')\ssbs(\alpha,a)\rbr=\beta\ou\non\alpha\ou\alpha'=1$.

\smallskip\noindent
It remains to show that $A'$ has no minimal $\veps$-element for $\sbs_{\cal D}$ and for $<_{\non\alpha'}$.\\
Therefore, let $u\eps A'$, thus $u=\alpha' x\sqcup(\non\alpha)y\sqcup\beta z$ with $x,y\eps A$ and $z\eps B$.\\
By hypothesis on $A,B$, there exists $x',y'\eps A,x'\sbs_{\cal D}x,y'<_{\non\alpha}y$ and
$z'\eps B,z'<_{\beta}z$.\\
Then, if we set \ $u'=\alpha'x'\sqcup(\non\alpha)y'\sqcup\beta z'$, we have $u'\eps A'$.\\
Moreover, we have $\lbr u'\sbs u\rbr\ge\alpha'\et\lbr x'\sbs x\rbr$, and therefore ${\cal D}\lbr u'\sbs u\rbr$,
that is $u'\sbs_{\cal D}u$.\\
Finally, $\lbr u'< u\rbr\ge(\non\alpha\et\lbr y'<y\rbr)\ou(\beta\et\lbr z'<z\rbr)
=\non\alpha\ou\beta=\non\alpha'$~; therefore, we have $u'<_{\non\alpha'}u$.

\qed

\begin{theorem}\label{MD_bien_fonde}
${\cal M}_{\cal D}$ is well founded, and therefore has the same ordinals as ${\cal N}'_\in$.
\end{theorem}\noindent
We apply theorem~\ref{sbs_D} to the binary relation $\in$ which is well founded in ${\cal M}$.
We deduce that the relation \ ${\cal D}\lbr x\in y\rbr$, that is \ $x\in_{\cal D}y$, is well founded in
${\cal N}$.

\qed

\smallskip\noindent
The relation $\in_{\cal D}$ is well founded \emph{and extensional}, which means that we have, in ${\cal N}$~:\\
\centerline{$\pt x\pt y\left(\pt z(z\in_{\cal D}x\dbfl z\in_{\cal D}y)\to
\pt z(x\in_{\cal D}z\to y\in_{\cal D}z)\right)$.}

\noindent
It follows that we can define a \emph{collapsing}, by means of a function symbol $\Phi$, which is an
\emph{isomorphism} of ${({\cal M}_{\cal D},\in_{\cal D})}$ on a  transitive class in the model
${\cal N}_\in$ of ZF\/, which contains the ordinals. This means that we have~:\\
\centerline{$\pt x\pt y(y\in_{\cal D}x\to\Phi(y)\in\Phi(x))$~; \
$\pt x(\pt z\in\Phi(x))(\ex y\in_{\cal D}x)\,z\simeq\Phi(y)$.}

\smallskip\noindent
The definition of $\,\Phi$ is analogous with that of the \emph{rank function} already defined for a
\emph{transitive} well founded relation.
The details will be given in a later version of this paper.\\
Il follows that~:
\begin{theorem}
The realizability model ${\cal N}_\in$ contains a transitive  class, which contains the
ordinals and is an  elementary extension of the ground model ${\cal M}$.
\end{theorem}\noindent

\begin{corollary}
The class \ $L^{\cal M}$ of constructible sets in ${\cal M}$ is an elementary submodel of~$\,L^{\cal N}$.
\end{corollary}

\end{document}